\documentclass[prd,superscriptaddress,amsfonts,amssymb,amsmath,showpacs,twocolumn,nofootinbib]{revtex4-2}
\usepackage[utf8]{inputenc}
\usepackage{graphicx}
\usepackage{mathtools}
\usepackage[margin=0.75in]{geometry}
\usepackage{times}
\usepackage{microtype}
\usepackage{setspace}
\usepackage{parskip}
\usepackage{hyperref}
\usepackage{float} 
\usepackage{xcolor}
\usepackage{booktabs}
\usepackage{color}

\usepackage{amsfonts}
\usepackage{latexsym}
\usepackage{graphicx}
\usepackage{amsmath}
\usepackage{palatino}
\usepackage{mathpazo}
\usepackage{adjustbox}
\usepackage{tensor}
\usepackage{textcomp}
\usepackage{float}
\usepackage{booktabs}
\usepackage{dcolumn}
\usepackage[titletoc]{appendix}
\usepackage{booktabs}
\usepackage{multirow}
\usepackage{hyperref}
\hypersetup{colorlinks,citecolor=blue}
\usepackage{amsmath}
\usepackage{xcolor}
\usepackage{epsfig}
\usepackage{caption}
\usepackage{subcaption}
\hypersetup{colorlinks,citecolor=blue}
\hypersetup{colorlinks=true,linkcolor=magenta,filecolor=magenta,    urlcolor=blue}

\begin{document}

\title{Probing the Evolution of Dark Energy: A Joint Analysis of DESI DR2, Pantheon+, and Cosmic Chronometers}
\author{Chanchal Kumari}
\email{chanchalphy18@gmail.com}
\affiliation{Department of Physics, University of Rajasthan, Jaipur 302004, India}

\author{Dinesh Kumar}
\email{dineshsuman09@gmail.com}
\affiliation{Department of Physics, University of Rajasthan, Jaipur 302004, India}
\preprint{}
\begin{abstract}
    We investigate several phenomenological dark energy parameterizations using a joint analysis of late-time cosmological observations, including cosmic-chromatometer measurements of the Hubble parameter, DESI DR2 baryon acoustic oscillation data, and the Pantheon+ Type Ia supernova sample. Our results show that allowing for a time-varying dark energy equation of state significantly improves the overall fit compared to $\Lambda$CDM. The present-day equation-of-state parameter departs from the standard cosmological constant value. In contrast, the evolution parameter in two-parameter models tends to be negative, indicating a possible time dependence of dark energy. However, the constraints on the evolution remain moderate, and current data cannot clearly distinguish the specific functional form of dark energy. Model comparison using information criteria suggests that dynamical dark energy models are favored over $\Lambda$CDM, with the most straightforward one-parameter extension emerging as the most parsimonious scenario. These findings indicate a mild preference for dark energy evolution, though future high-precision observations will be required for definitive conclusions.
\end{abstract}

\maketitle

\section{Introduction}
The observational evidence of the accelerated expansion of the universe was first revealed through luminosity–distance measurements of distant Type Ia supernovae (SNe Ia) \cite{SupernovaSearchTeam:1998fmf, SupernovaCosmologyProject:1998vns}, which had a profound impact on modern cosmology. Subsequent and independent measurements, including baryon acoustic oscillations (BAO) \cite{SDSS:2005xqv, Percival:2007yw} and precise measurements of the cosmic microwave background (CMB)\cite{WMAP:2003elm,WMAP:2003ivt,Planck:2018vyg}, have further supported the accelerated universe. The standard model of cosmology, known as $\Lambda CDM$, describes the present universe, which is composed of baryons, cold dark matter, and dark energy in the form of cosmological constant $\Lambda$.

Despite its remarkable success, the $\Lambda$CDM model faces some theoretical and observational challenges. On the theoretical side, the discrepancy between the observational cosmological constant $\Lambda$ and the predictions of quantum field theory is known as the cosmological constant problem \cite{Weinberg:1988cp}. On the observational side, persistent tension between the measurements of the present-day expansion rate of the universe obtained from local observations and those inferred from early-universe data. In particular, analyses based on Cepheid calibrated Type Ia supernova provide a local estimate of the Hubble constant of $H_0 = 73.04 \pm 1.04 \ \mathrm{km\,s^{-1}\, Mpc^{-1}}$ \cite{Riess:2021jrx}, whereas measurements from Cosmic Microwave background (CMB) observations by the Planck mission yield a lower value of $H_0 = 67.4 \pm 0.5 \ \mathrm{km\,s^{-1}\, Mpc^{-1}}$ \cite{Planck:2018vyg}. The difference between these determinations corresponds to a statistical tension ranging from $\sim 4\sigma$ to $6\sigma$, depending on analysis method and dataset used (see refs. \cite{FernandezArenas:2017dux, Riess:2019cxk, H0LiCOW:2019pvv, Freedman:2019jwv, Yuan:2019npk, Huang:2019yhh, Pesce:2020xfe, deJaeger:2020zpb, Schombert:2020pxm, Blakeslee:2021rqi, Montani:2025nmz}). Recent studies suggest that the disagreement in the Hubble constant has become more significant. A joint analysis of CMB data from Planck, SPT-3G, and ACT \cite{Planck:2018vyg, Planck:2018nkj, ACT:2020gnv,AtacamaCosmologyTelescope:2025blo,SPT-3G:2025bzu}, when compared with a uniformly compiled set of late-universe measurements from the $H_0$ Distance Network (H0DN), indicates a tension at the level of roughly $7.1\sigma$ \cite{H0DN:2025lyy}. This currently stands as the most compelling evidence for the $H_0$ discrepancy. This persistent discrepancy has become one of the most compelling challenges in cosmology (for reviews, see refs. \cite{Verde:2019ivm, Riess:2019qba, DiValentino:2021izs, Perivolaropoulos:2021jda, Abdalla:2022yfr, DiValentino:2022fjm, Hu:2023jqc})  and has motivated the exploration of new physics beyond the standard cosmological model.

Although $\Lambda$CDM provides an excellent fit to a variety of cosmological data, it is widely considered as an approximate description of a more fundamental underlying theory that remains to be discovered. Several extensions have been explored using multiple observational probes \cite{Knox:2019rjx, DiValentino:2020zio, Kamionkowski:2022pkx, Khalife:2023qbu}. Within the framework of the General Theory of Relativity, the simplest extension of $\Lambda$CDM arises by allowing dark energy to be dynamical rather than a cosmological constant value $-1$. In such scenarios, the equation-of-state parameter $w_{DE}= p_{DE}/\rho_{DE}$ can take either a constant value different from $-1$ or can be a function of redshift $z$. The literature contains a wide variety of parameterization describing the dark energy equation-of-state parameter $w_{DE}$ \cite{Astier:2000as, Chevallier:2000qy, Linder:2002et,   Weller:2001gf, Jassal:2005qc, Linder:2005ne, Gong:2005de, Nesseris:2005ur, Feng:2004ff, Basilakos:2006us, Nojiri:2006ww, Nojiri:2006ww, Saridakis:2008fy, Barboza:2008rh, Saridakis:2009pj, Dutta:2009yb, Saridakis:2009ej, Ma:2011nc, Feng:2011zzo, Feng:2012gf, DeFelice:2012vd, Chen:2011cy, Basilakos:2013vya, Umilta:2015cta,  DiValentino:2016hlg, DiValentino:2017zyq, DiValentino:2017gzb, Zhao:2017cud, DiValentino:2017rcr, Marcondes:2017vjw, Yang:2017alx, Pan:2017zoh, Li:2024qus}.

BAO measurements reported since early 2024 from the first data release of the Dark Energy Spectroscopic Instrument (DESI) \cite{DESI:2024mwx} have posed significant challenges to the $\Lambda$CDM framework, and these tensions have further intensified with the subsequent DR2 data \cite{DESI:2025zgx}. The DESI team has provided a model-independent analysis showing a strong hint of dynamical dark energy, indicating that the equation-of-state parameter $w_{DE}$ is not constant.

In this paper, we investigate a class of dynamical dark energy models characterized by different parameterizations of the equation-of-state parameter $w_{DE}$, motivated by recent observational developments that challenge the $\Lambda$CDM framework. To capture possible time evolution in dark energy, we consider several widely studied parameterizations, including the $w_0$CDM, Chevallier–Polarski–Linder (CPL), Barboza-Alcaniz (BA), Jassal–Bagla–Padmanabhan (JBP), logarithmic, and exponential. 
By exploring a suite of these phenomenological dark energy parameterizations, we aim to determine whether the current data necessitates a departure from $w = -1$ paradigm. Our goal is to assess whether a time-varying equation of state offers a more statistically parsimonious explanation for the observed expansion rate, thereby providing clues toward a more fundamental theory of the dark sector.

These models are constrained by a combination of observational datasets that probe the universe's expansion history across different redshift ranges. In particular, we include Hubble parameter measurements from cosmic chronometers, DESI BAO measurements, and the latest type Ia supernova compilation from the Pantheon+ dataset \cite{Planck:2018vyg, Brout:2022vxf}. To estimate the model parameters, we adopt a Markov Chain Monte Carlo (MCMC) approach  \cite{Salvati:2020cluster}, which efficiently explores the multidimensional parameter space and yields reliable posterior distributions for the cosmological parameters. Furthermore, to assess the relative performance of the considered models with respect to the $\Lambda$CDM scenario, we employ model selection criteria such as the Akaike Information Criterion (AIC) and the Bayesian Information Criterion (BIC) \cite{Wang:2024carbon}. These statistical tools enable us to quantify whether the introduction of additional degrees of freedom associated with dynamical dark energy is justified by the improvement in the fit to observational data.

This paper is organized as follows. In Section \ref{models}, we describe the theoretical framework and dark energy parameterizations used for this analysis. Section \ref{data} provides details on the datasets used in our analysis and also discuss about the analysis methodology. Section \ref{Results} describes the findings of our analysis. We conclude our paper in Section \ref{conclusions}.
 
\section{Theoretical Framework}
\label{models}

We assume a spatially flat Friedmann–Lemaitre–Robertson–Walker (FLRW) Universe described by the metric \cite{Ryden}
\begin{equation}
ds^2 = -c^2 dt^2 + a^2(t)\left[dr^2 + r^2 (d\theta^2+\sin^2\theta d\phi^2)\right],
\end{equation}
where the scale factor $a(t)$ represents the expansion of the universe, and this factor is normalized such that $a(t_0) = 1 $ at the present epoch. Observations of the CMB and large-scale structure provide strong support for a spatially flat Universe \cite{Planck:2018vyg}.

The Friedmann equation governs the expansion dynamics
\begin{equation}
\frac{H^2(z)}{H_0^2} = \Omega_m (1+z)^3 + \Omega_r (1+z)^4 + (1-\Omega_{\rm m} - \Omega_r)f_{de}(z)
\end{equation}
where $H(z)$ is the Hubble parameter, $H_0$ its present value, and $\Omega_i$ denote present-day density parameters. Since our analysis focuses on late-time cosmology, the radiation component is negligible and is omitted in the numerical analysis.

The dark energy density evolution is determined by its equation-of-state parameter
\begin{equation}
w(z) = \frac{p_{\rm DE}}{\rho_{\rm DE}},
\end{equation}
which may be constant or time-dependent. For a general $w(z)$, the dark energy density evolves as
\begin{equation}
\rho_{\rm DE}(z) = \rho_{\rm DE,0} f_{de}(z)
\end{equation}
where 
\begin{equation}
f_{de}(z) = \exp \left[ 3 \int_0^z \frac{1+w(z')}{1+z'} dz' \right]
\label{fde}
\end{equation}

In this work, we consider several phenomenological parameterizations of a time-varying dark energy equation of state, as summarized in Table \ref{tab:de_models}. These functional forms provide a model-independent framework to capture potential deviations from the $\Lambda$CDM baseline $w = -1$. By substituting these definitions into Eq. \ref{fde}, we derive the corresponding energy density evolution $f_{de}(z)$, which uniquely determines the expansion history $H(z)$. This allows us to perform a robust Bayesian likelihood analysis against the combined late-time datasets.

\begin{table*}[t]
\centering
\caption{Dark energy parameterizations considered in this work.}
\label{tab:de_models}
\renewcommand{\arraystretch}{1.2}
\begin{tabular}{lcc}
\hline
\textbf{Model} & \textbf{Equation of State $w(z)$} & \textbf{Free Parameters} \\
\hline

$\Lambda$CDM 
& $w = -1$ 
& $\Omega_m, H_0$ \\

$w_0$CDM 
& $w = w_0$ 
& $\Omega_m, H_0, w_0$ \\

CPL ($w_0w_1$CDM) 
& $w(z) = w_0 + w_1 \frac{z}{1+z}$ 
& $\Omega_m, H_0, w_0, w_1$ \\

Barboza--Alcaniz (BA) 
& $w(z) = w_0 + w_1 \frac{z(1+z)}{1+z^2}$ 
& $\Omega_m, H_0, w_0, w_1$ \\

Jassal--Bagla--Padmanabhan (JBP) 
& $w(z) = w_0 + w_1 \frac{z}{(1+z)^2}$ 
& $\Omega_m, H_0, w_0, w_1$ \\

Logarithmic 
& $w(z) = w_0 + w_1 \ln(1+z)$ 
& $\Omega_m, H_0, w_0, w_1$ \\

Exponential 
& $w(z) = w_0 + w_1 \left(e^{\frac{z}{1+z}} - 1\right)$ 
& $\Omega_m, H_0, w_0, w_1$ \\

\hline
\end{tabular}
\end{table*}

\section{Observational Data and Methodology}
\label{data}
In this section, we describe the observational datasets used in our analysis and outline the statistical methodology for exploring the parameter space of cosmological dark energy models. In this study, we used the data from cosmic chronometers, baryon acoustic oscillations from DESI Data Release 2, and Type Ia supernovae from the Pantheon+ compilation.

\subsection{Cosmic Chronometers}
Galaxies exhibiting passive stellar evolution on timescales much longer than their relative age differences can serve as cosmic chronometers, allowing a direct determination of the Hubble parameter using the differential-age method within an FLRW cosmology. The expansion rate is related to the differential age evolution through \cite{Jimenez:2001gg}
\begin{equation}
     H(z)=-\frac{1}{1+z} \frac{dz}{dt} 
\end{equation}

We use the 32 cosmic chronometer data points in redshift range of $0.07<z<1.965$ \cite{Moresco:2016mzx}. Cosmic chronometer observations offer a largely model-independent way to trace the Universe’s expansion history by directly measuring the differential aging of passively evolving galaxies.

To test the viability of cosmological models against observational datasets, we use the chi-squared ($\chi^2$) statistic. The model parameters that best describe the observational datasets are obtained by minimizing the $\chi^2$ function. To obtain the parameter space, the $\chi^2$ function for CC data is defined as
\begin{equation}
\chi^2_{CC} =
\sum_{i=1}^{N}
\frac{[H_{theo}(z_i) - H_{obs}(z_i)]^2}
{\sigma_{H,i}^2}.
\end{equation}

\subsection{Baryon Acoustic Oscillation}
We have used the latest BAO measurements released by the DESI collaboration \cite{DESI:2025zgx}. A variety of tracers, including quasars (QSO), Lyman-$\alpha$ forest quasars (Ly$\alpha$ QSO), the Bright Galaxy Survey (BGS), luminous red galaxies (LRG), emission-line galaxies (ELG), and a combined LRG+ELG sample, are used to extract these observations. The BAO measurements are interpreted using three distance measures: the transverse comoving distance ($D_M$), the line-of-sight distance ($D_H$), and the angle-averaged distance ($D_V$). These BAO observables are defined as: 
\begin{align}
\tilde{D}_M(z) = \frac{D_{M}(z)}{r_d} &= \frac{c}{r_d} \int_{0}^{z} \frac{dz'}{H(z')},\\
\tilde{D}_H(z) = \frac{D_{H}(z)}{r_d} &= \frac{c}{r_d \, H(z)},\\
\tilde{D}_V(z) = \frac{D_{V}(z)}{r_d} &= \frac{\left[ z \, D_{H}(z) \, D_{M}^{2}(z) \right]^{1/3}}{r_d}.
     \end{align}
where $r_d$ is the sound horizon at the drag epoch. We consider $r_d = 147.09 \pm 0.20 $ Mpc \cite{ref35} in our analysis. We defined the chi square function for these data as:
\begin{equation}
\chi^2_{BAO} = \chi^2_{\tilde{D}_M} + \chi^2_{\tilde{D}_H} + \chi^2_{\tilde{D}_V}
\end{equation}
where
\begin{align}
\chi^2_{\tilde{D}_M} &= \sum \frac{[\tilde{D}_M^{theo}(z_i) - \tilde{D}_M^{exp}(z_i)]^2}
{\sigma_{i,\tilde{D}_M}^2}\\
\chi^2_{\tilde{D}_H} &= \sum \frac{[\tilde{D}_H^{theo}(z_i) - \tilde{D}_H^{exp}(z_i)]^2}
{\sigma_{i,\tilde{D}_H}^2} \\
\chi^2_{\tilde{D}_V} &= \sum \frac{[\tilde{D}_V^{theo}(z_i) - \tilde{D}_V^{exp}(z_i)]^2}
{\sigma_{i,\tilde{D}_V}^2}
\end{align}

\subsection{Supernovae of Type Ia}
We have used the Pantheon+ dataset, which contains 1701 light curves from 1550 spectroscopically confirmed Type Ia supernovae (SNe Ia) covering the redshift range $0.00122 \leq z \leq 2.2613$ \cite{Scolnic:2021amr}. The full dataset and covariance matrix for Pantheon + are publicly available via GitHub repository \footnote{\url{https://github.com/PantheonPlusSH0ES/DataRelease/tree/main/Pantheon\%2B_Data}}. For Type Ia supernovae, the observed quantity is the rest-frame apparent magnitude $m_B$
, which encodes information about both the intrinsic brightness $M_B$ and the luminosity distance $d_L$
\begin{equation}
 m_B = M_B + 5 \log_{10} \left( d_{L}(z)/Mpc \right) + 25
\end{equation}

The distance modulus $\mu = m_B - M$ is observed experimentally, so we compare the theoretical distance modulus in our analysis. We define the difference between the theoretical and observed distance modulus as
\begin{equation}
\Delta \mu = \mu_{exp} - \mu_{theo}
\end{equation}

The chi-squared function for the Type Ia supernovae dataset is given by
\begin{equation}
\chi^2_{SN} = A - \frac{B^2}{D} + ln\, \frac{D}{2\pi},
\end{equation}
where $A = \Delta \mu^T \cdot \mathcal{C}^{-1}  \cdot \Delta \mu, B = \Delta \mu^T \cdot \mathcal{C}^{-1}  \cdot 1$ and $D = 1^T \cdot \mathcal{C}^{-1}  \cdot 1$, $\mathcal{C}$ is the covariance matrix provided by Pantheon+ team which includes the statistical and systematic uncertainties. 

Based on the observational datasets, the total chi-squared function can be expressed as
\begin{equation}
\chi^2_{total} = \chi^2_{CC} + \chi^2_{BAO} + \chi^2_{SN}
\end{equation}
To estimate the best-fit values of our model parameters, we use the Markov Chain Monte Carlo (MCMC) method implemented in the open-source Python package {\it emcee} \cite{Foreman-Mackey:2012any}.

The relative performance of the different dark energy models is assessed using AIC and BIC, which provide quantitative measures of goodness-of-fit while penalizing model complexity. The AIC and BIC are defined as
\begin{align}
AIC &= -2 \, ln \, \mathcal{L}_{max} + 2n \\
BIC &= -2 \, ln \, \mathcal{L}_{max} + n \, ln \, N
\end{align}

where $ln \, \mathcal{L}_{max}$ represents the maximum log-likelihood, n denotes the number of free parameters in the model, and N is the number of data points. We compare the models with respect to the $\Lambda$CDM, defining $\Delta X = X_{\Lambda CDM} - X_i$, where $X = AIC, BIC$ and $i$ denotes the different dark energy models. 

\section{Results and Discussion}
\label{Results}
\begin{figure*}[!htb]
\centering

\begin{subfigure}{0.33\textwidth}
    \includegraphics[width=\linewidth]{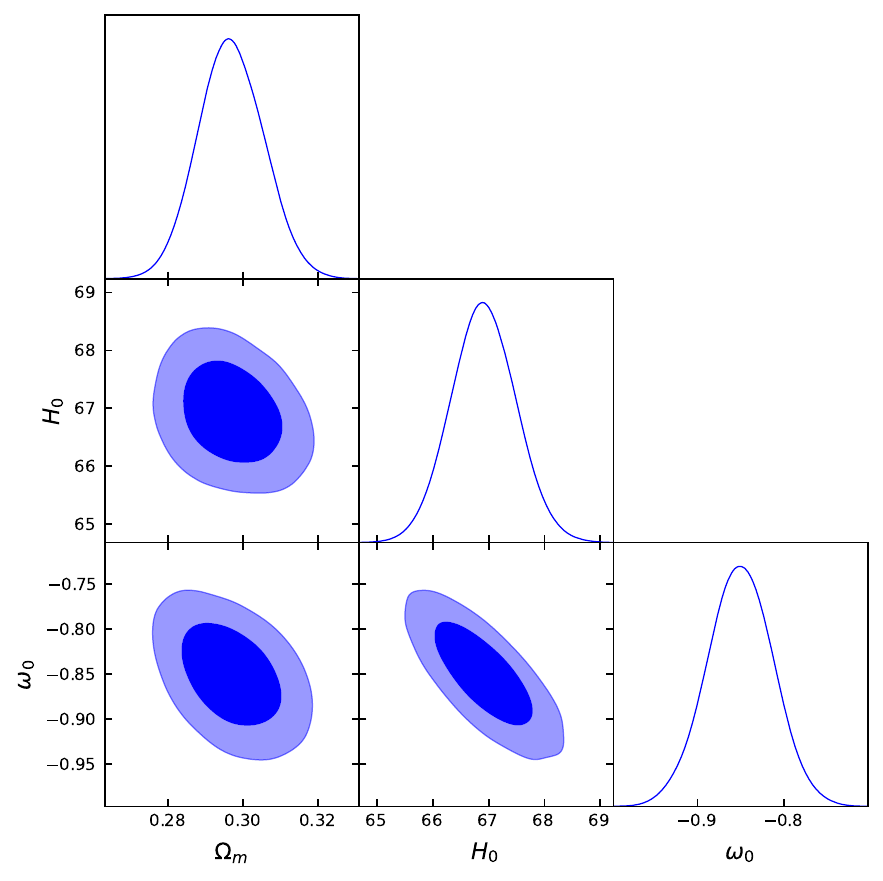}
    \caption{$\omega_0$CDM Model}
    \label{fig_1a}
\end{subfigure}\hfill
\begin{subfigure}{0.33\textwidth}
    \includegraphics[width=\linewidth]{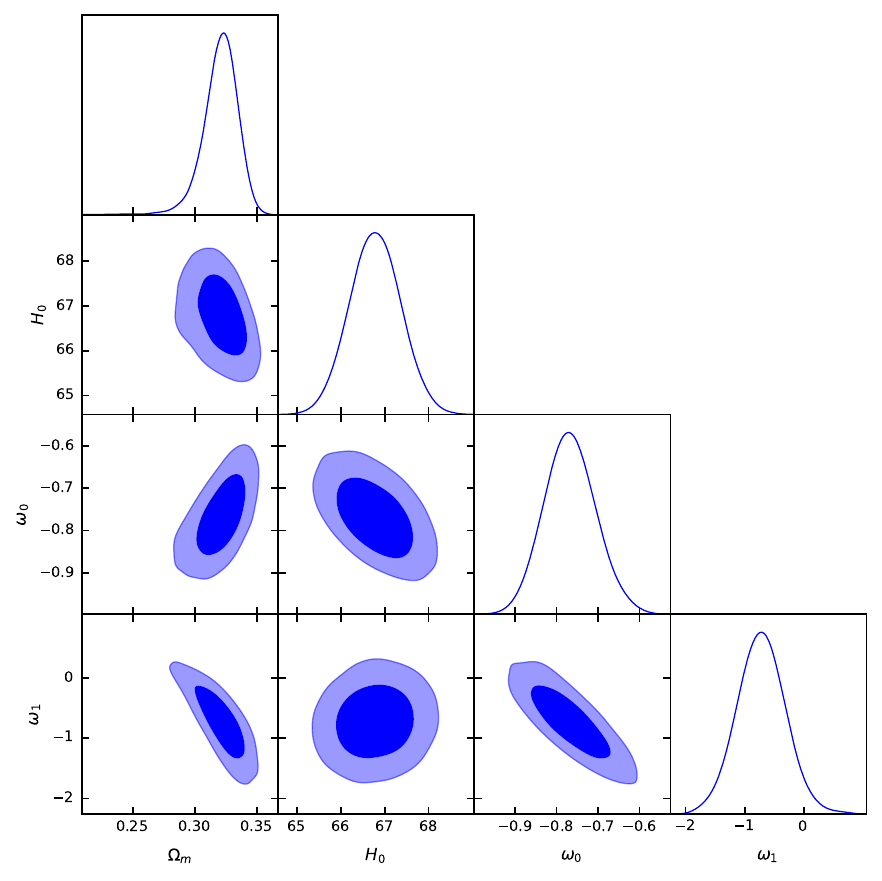}
    \caption{CPL Model}
    \label{fig_1b}
\end{subfigure}\hfill
\begin{subfigure}{0.33\textwidth}
    \includegraphics[width=\linewidth]{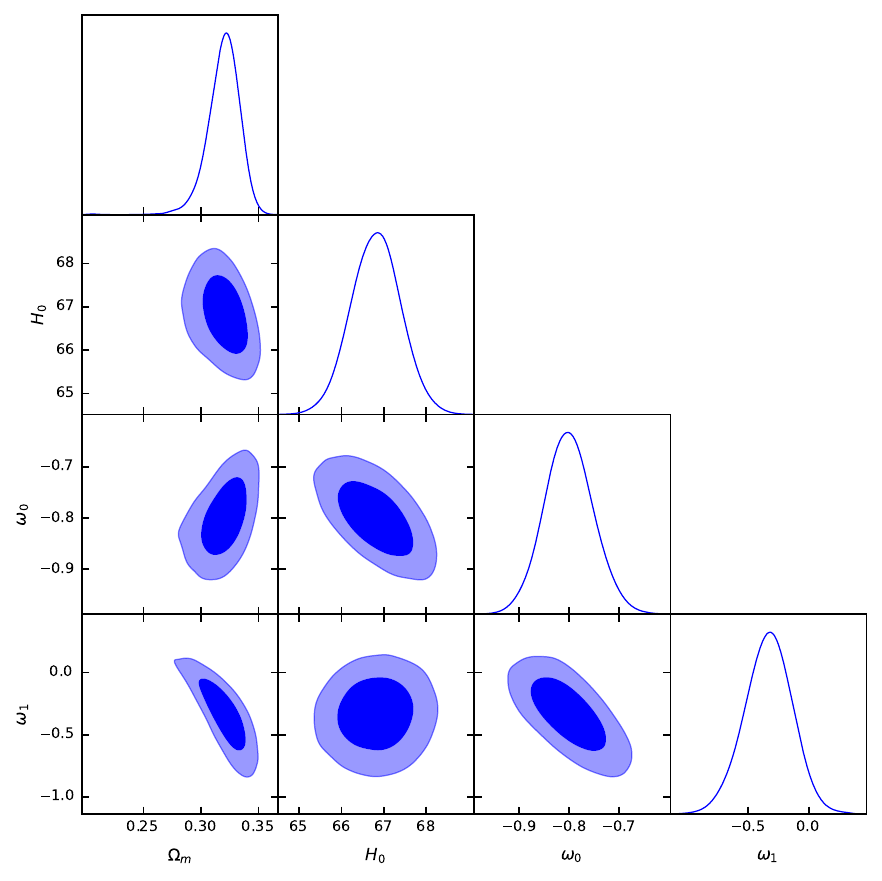}
    \caption{BA Parameterization}
    \label{fig_1c}
\end{subfigure}

\par\medskip 

\begin{subfigure}{0.33\textwidth}
    \includegraphics[width=\linewidth]{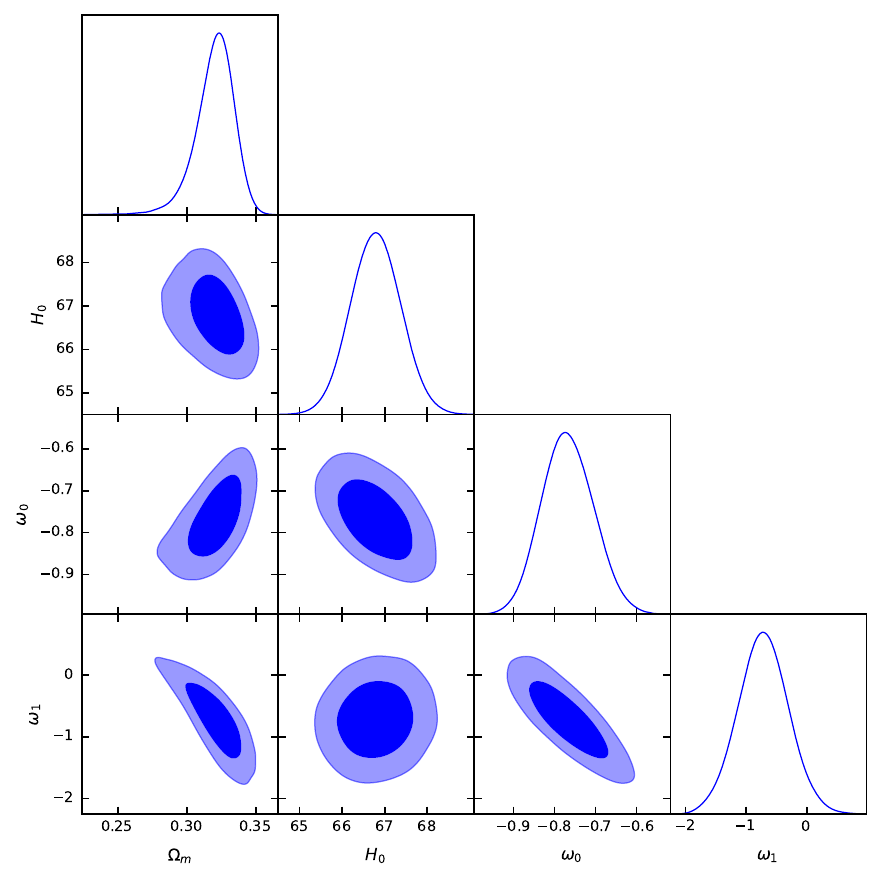}
    \caption{JBP Parameterization}
    \label{fig_1d}
\end{subfigure}\hfill
\begin{subfigure}{0.33\textwidth}
    \includegraphics[width=\linewidth]{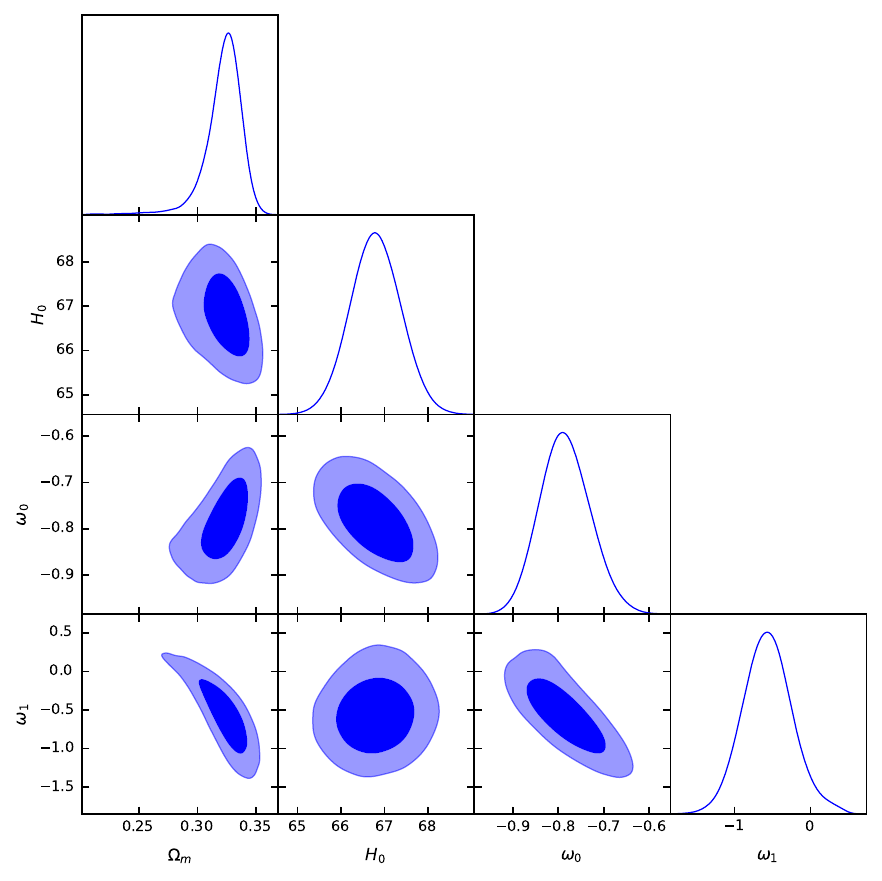}
    \caption{Logarithmic Parameterization}
    \label{fig_1e}
\end{subfigure}\hfill
\begin{subfigure}{0.33\textwidth}
    \includegraphics[width=\linewidth]{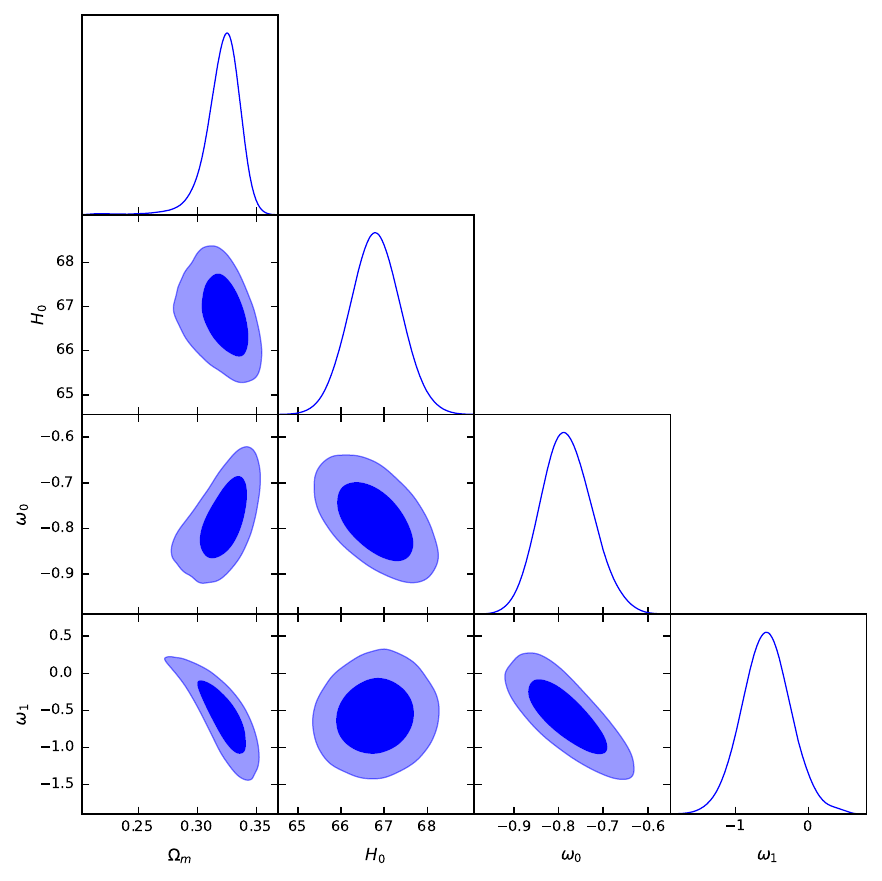}
    \caption{Exponential Parameterization}
    \label{fig_1f}
\end{subfigure}

\caption{One-dimensional posterior distributions and two-dimensional confidence contours at $68\%$ and $95\%$ CL for $\omega_0$CDM, CPL, BA, JBP, logarithmic, and exponential parameterizations using CC, DESI BAO, and Pantheon+ datasets.}
\label{fig_cc}
\end{figure*}

In this section, we outline and interpret the findings obtained from our analysis. We analyzed the cosmological models using a combination of current observational datasets. The dataset includes the Hubble constant from the cosmic chronometer, DESI BAO, and Pantheon+. The constraints on the model parameters are summarized in the table \ref{result}. One-dimensional posterior distribution and the marginalized constraints with $68\%$ and $95\%$ confidence contours are shown in figures \ref{fig_cc}.

\begin{table*}[!htb]
    \centering
    \caption{Best-fit values of cosmological parameters (with 1$\sigma$ asymmetric uncertainties) for dark energy models.}
    \label{tab:params_transposed}
    \renewcommand{\arraystretch}{1.3} 
    \setlength{\tabcolsep}{6pt}
   
    \resizebox{\textwidth}{!}{
    \begin{tabular}{lcccccccc}
        \hline
        \textbf{Parameter} & \textbf{$\Lambda$CDM} & \textbf{$w_o$CDM} & \textbf{CPL}  & \textbf{BA} & \textbf{JBP} & \textbf{Logarithmic} & \textbf{Exponential} \\
        \hline
        $H_0$         & $68.48^{+0.46}_{-0.46}$ & $66.91^{+0.58}_{-0.58}$ & $66.78^{+0.58}_{-0.58}$  & $66.82^{+0.57}_{-0.59}$ & $66.79^{+0.59}_{-0.58}$ & $66.79^{+0.59}_{-0.57}$ & $66.80^{+0.58}_{-0.58}$ \\
        \hline
        $\Omega_m$          & $0.31^{+0.01}_{-0.01}$ & $0.30^{+0.01}_{-0.01}$ & $0.32^{+0.01}_{-0.01}$  & $0.32^{+0.01}_{-0.01}$ & $0.32^{+0.01}_{-0.01}$ & $0.32^{+0.01}_{-0.02}$ & $0.32^{+0.01}_{-0.02}$ \\
        \hline
        $w_0$               & -- & $-0.85^{+0.04}_{-0.04}$ & $-0.77^{+0.06}_{-0.06}$ & $-0.80^{+0.05}_{-0.05}$ & $-0.77^{+0.07}_{-0.06}$ & $-0.79^{+0.06}_{-0.05}$ & $-0.78^{+0.06}_{-0.06}$ \\
        \hline
        $w_1$               & -- & -- & $-0.72^{+0.40}_{-0.40}$ & $-0.33^{+0.19}_{-0.20}$ & $-0.72^{+0.41}_{-0.41}$ & $-0.57^{+0.33}_{-0.32}$ & $-0.58^{+0.33}_{-0.33}$ \\
        \hline
        $\chi^2$ & $1802.76$ & $1787.88$ & $1784.60$ & $1784.79$ & $1784.60$ & $1784.56$ & $1784.60$ \\
        \hline
        $\Delta$ AIC & $0$ & $12.88$ & $14.16$ & $13.97$ & $14.16$ & $14.20$ & $14.16$ \\
        \hline
         $\Delta$ BIC & $0$ & $7.4$ & $3.22$ & $3.03$ & $3.22$ & $3.25$ & $3.22$ \\
        \hline
    \end{tabular}
    \label{result}
    }
\end{table*}

The standard $\Lambda$CDM model provides a minimum $\chi^2 = 1802.76$, with best-fit values $H_0 = 68.48 \pm 0.46$~km~s$^{-1}$~Mpc$^{-1}$ and $\Omega_m = 0.31 \pm 0.01$. When dark energy dynamics are allowed, all extended models lead to a significant improvement in the fit, reducing the minimum $\chi^2$ by $\Delta \chi^2 > 15$ relative to $\Lambda$CDM. This indicates that the combined late-time data favor models beyond $\Lambda$CDM.

In the $w_0$CDM scenario, the dark energy equation-of-state parameter is constrained to $w_0 = -0.85 \pm 0.04$, which indicates a $\sim4\sigma$ deviation from the cosmological constant value $w = -1$ with the datasets taken in this analysis. Due to the additional degree of freedom in $w_0$CDM, the Hubble parameter value reduces to $H_0 = 66.91 \pm {0.58}$ in compare to $H_0 = 68.48 \pm {0.46}$ in $\Lambda$CDM. The $\Delta \chi^2  = \chi^2_{model}-\chi^2_{\Lambda CDM} > 15 $ indicates that all models provide a better fit to the data over the $\Lambda CDM$ model. A lower $\chi^2$ value of all dynamical dark energy models suggests that deviations from the cosmological constant from $w = -1$ improve the agreement of data. However, the $\chi^2$ values for all models beyond $\Lambda CDM$ are almost the same, which implies that based on $\chi^2$ values, no strong statistical preference can be established for one dynamical dark matter parameterization over another.

For the two-parameter models, including the CPL ($w_0w_1$), BA, JBP, logarithmic, and exponential parameterizations, the present-day equation-of-state parameter lies in the range $-0.80 \lesssim w_0 \lesssim -0.77$. This indicates a mild deviation from the cosmological constant value $w = -1$. The evolution parameter $w_1$ is found to be negative in all cases, with typical values in the range $-0.7 \lesssim w_1 \lesssim -0.3$, although the associated uncertainties remain relatively large. This reflects the limited sensitivity of the current data to the detailed time evolution of dark energy. The evolution parameter $w_1$ is found to be non-zero at $>1.5\sigma$ for all these models.

According to the Akaike Information Criterion, the dynamical dark energy model parameterization is strongly favored over $\Lambda CDM$ with $\Delta AIC > 10$. The additional degree of freedom introduced in dynamical dark energy models improves the goodness of fit, making these models preferable to $\Lambda CDM$. The BIC applies a stronger penalty for additional parameters. The $\Delta BIC$ value indicates the preference of dynamical dark energy models over the $\Lambda CDM$. In particular, the one-parameter $w_0 CDM$ model is strongly favored while the two-parameter models are statistically competitive with smaller $\Delta BIC$ values.

Overall, the combined statistical analyses indicate a preference for dynamical dark energy models over $\Lambda CDM$. The one-parameter model $w_0 CDM$ emerges as the most economical and statistically supported model with the observational constraints.

\section{Summary and Conclusions}
\label{conclusions}

In this work, we have investigated a set of dynamical dark energy models using a combined analysis of cosmic chronometer, BAO, and Type Ia supernova data. In addition to the standard $\Lambda$CDM scenario, we considered several commonly used parameterizations of an evolving dark energy equation of state, including the $w_0$CDM, CPL, BA, JBP, logarithmic, and exponential.

Our results show that dynamical dark energy models provide an improvement in the goodness of fit compared to the $\Lambda$CDM model. All dynamical dark energy models reduce the minimum $\chi^2$ by approximately $\Delta\chi^2 > 15$ relative to $\Lambda$CDM when fitting the combined CCH+BAO+SNe Ia dataset. All dynamical dark energy models yield nearly identical $\chi^2$ values, suggesting that current data do not strongly distinguish the functional form of dark energy evolution. The improvement in $\chi^2$ values is also reflected in the Akaike Information Criterion, which favors the extended models.

We find that in one EOS parameter model $w_0 CDM$, the equation of state parameter is constrained to $w_0 = 0.85 \pm 0.04$, which deviates from the cosmological constant value $w = -1$. The Hubble parameter value in this model is slightly lower than the $\Lambda CDM$. For the two-parameter models, the present-day value $w_0$ provides a deviation from $w =-1$, while the evolution parameter $w_1$ is consistently negative with large uncertainties for all the parameterizations. This suggests a mild indication of evolving dark energy, but the time dependence is weakly constrained.

The information criteria AIC and BIC further clarify the evidence of dark energy parameterization beyond $\Lambda$CDM. The AIC value shows a strong preference for dynamical dark energy models over $\Lambda$CDM, indicating that the improvement in fit compensates for the additional degrees of freedom. The Bayesian Information Criterion, which penalizes model complexity more strongly, favors the simpler $w_0$CDM extension, while the two-parameter models remain statistically competitive but less strongly supported.

Overall, our analysis indicates that late-time cosmological observational data prefer the dark energy models with dynamical behavior over the $\Lambda$CDM scenario. In contrast, a specific functional form of dark energy equation of state cannot be distinguished with the datasets considered in this analysis. The $w_0$CDM model emerges as the most economical and statistically supported model. Future high-precision surveys will be essential for constraining dark energy dynamics and assessing whether deviations from a cosmological constant are statistically significant.

\section*{Acknowledgement}
DK acknowledges support from the ANRF, Government of India, under research grant no. SERB/EEQ/2021/000965.

\end{document}